\documentclass[a4paper,11pt]{article}
\pdfoutput=1 

\usepackage{jheppub} 

\usepackage[T1]{fontenc} 
\usepackage{comment}
\usepackage{soul}

\usepackage[textsize=footnotesize,textwidth=2.cm]{todonotes}

\def\p{\partial}

\newcommand{\ba}{\begin{array}}
\newcommand{\ea}{\end{array}}
\newcommand{\bi}{\begin{itemize}}
\newcommand{\ei}{\end{itemize}}
\newcommand{\bea}{\begin{eqnarray}}
\newcommand{\eea}{\end{eqnarray}}
\newcommand{\be}{\begin{equation}}
\newcommand{\ee}{\end{equation}}
\newcommand{\nn}{\nonumber}

\title{\boldmath Gravitational radiations of Kerr black hole from warped symmetries}

\author{Jianfei Xu}
\date{\today}

\affiliation{Shing-Tung Yau Center and School of Mathematics, Southeast University, Nanjing, 210000, China}
\emailAdd{jfxu@seu.edu.cn}

\abstract{The warped conformal symmetries have been found in the covariant phase space of a set of non-trivial diffeomorphism near the Kerr black hole horizon. In this paper, we consider the retarded Green's function and the absorption probability for the scalar and higher spin perturbations on a generic non-extreme Kerr black hole background, and perform their holographic calculations in terms of the warped CFT. This provide further evidence on the conjecture that the warped CFT is relevant to the microstate description of the Kerr black hole.}

\begin{document}
\maketitle
\flushbottom
\section{Introduction}
In exploring the gravitational phenomena in the vicinity of a black hole, the conformal field theory (CFT) plays an essential role. This was not only due to the holographic duality (anti de-Sitter (AdS))$/$CFT, but also due to the large amount of studies go beyond it. One of the typical example is the Kerr$/$CFT~\cite{Guica:2008mu} which aims to understand the Kerr black hole in universe in terms of a two dimensional (2D) CFT. For an extreme Kerr black hole, there is a near horizon scaling region, known as the NHEK (near horizon extreme Kerr) geometry~\cite{Bardeen:1999px}, where the asymptotic symmetry analysis under specific boundary conditions shows that the dual field theory should be a chiral left-moving 2D CFT~\cite{Guica:2008mu}. Other boundary conditions that also implement the Kerr$/$CFT can be found in~\cite{Matsuo:2009sj, Chen:2011wt}. Evidence of the Kerr$/$CFT comes from the facts that the Cardy formula of a 2D CFT reproduces the black hole entropy~\cite{Guica:2008mu, Matsuo:2009sj} and the matching of CFT correlators with scattering amplitudes of perturbations on a near extreme Kerr black hole background~\cite{Bredberg:2009pv}.

For a non-extreme Kerr black hole, the NHEK geometry disappears and the near horizon geometry is just a Rindler space without conformal symmetry. However, as proposed in~\cite{Castro:2010fd}, a geometry with conformal symmetry is not a necessary condition for the interactions to exhibit conformal invariance. The conformal symmetries may hide in the solution space and still governs the dynamics of the fields. In the low frequency limit $\omega\ll1/M$, where $\omega$ is the frequency of the perturbative field and $M$ is the mass of the Kerr black hole, the spacetime can be divided into the near region $r\ll1/\omega$ and the far region $r\gg M$ which have a large overlap. The hidden conformal symmetry acting on the near region wave equation. Intuitively, this is highly because of the independence of choosing a matching surface in getting the full wave solution by the near-far matching~\cite{Castro:2010fd}. This hidden conformal symmetry can help to fix the black hole entropy and scattering amplitudes in term of the Cardy formula and correlators in a 2D CFT~\cite{Castro:2010fd}. The covariant phase space formalism~\cite{Haco:2018ske} and the gravitational wave analysis~\cite{Nian:2023dng} also support the validity of the hidden conformal symmetry.

In this paper, we will phenomenologically determine the scattering amplitudes of an non-extreme Kerr black hole by incident perturbative scalar and higher spin waves originating in the asymptotically flat region, and see whether or not the black hole will react like a warped CFT (WCFT) other than a 2D CFT. This will generalize the work in~\cite{Bredberg:2009pv} to the non-extreme case. The WCFT is a self-consistent field theory which also allows holographic interpretations. To be specific, it is a two dimensional non-relativistic quantum field theory with warped conformal symmetry. The symmetry algebra contains one Virasoro algebra and one $U(1)$ Kac-Moody algebra~\cite{Hofman:2011zj, Detournay:2012pc}. The global warped conformal symmetry is $SL(2, R)\times U(1)$. Concrete examples of WCFTs include chiral Liouville gravity~\cite{Compere:2013aya}, Weyl fermion models~\cite{Hofman:2014loa, Castro:2015uaa}, free scalars~\cite{Jensen:2017tnb}, and the complex Sachdev-Ye-Kitaev models as a symmetry-broken phase~\cite{Chaturvedi:2018uov}. Much like its CFT cousin, the two and three point correlation functions of the WCFT can be fixed by the global warped conformal symmetry up to normalizations and OPE coefficients~\cite{Song:2017czq}. The entanglement entropy and the outer-of-time-order correlators of the WCFT have been studied in~\cite{Castro:2015csg, Apolo:2018oqv}.

From a holographic perspective, WCFT was initiated from the studies of the holography of a large class of geometries with $SL(2, R)\times U(1)$ isometry. Typical examples are the warped AdS$_3$ (WAdS)~\cite{Anninos:2008fx, Compere:2008cv, Compere:2009zj, Blagojevic:2009ek, Anninos:2010pm, Anninos:2011vd, Henneaux:2011hv} and the NHEK geometry~\cite{Bardeen:1999px}. By imposing Dirichlet-Neumann boundary conditions on a AdS$_3$ in Einstein gravity, the WCFT can be viewed as an alternative holographic dual to the AdS$_3$~\cite{Compere:2013bya}. The corresponding conjectures WAdS$/$WCFT~\cite{Detournay:2012pc} and AdS$_3/$WCFT~\cite{Compere:2013bya} are possible holographic dualities beyond the standard AdS$/$CFT. Evidence of the above mentioned dualities is gathered from the following results. A Cardy like formula induced from the modular properties of WCFT partition functions recovers the thermal entropy of WAdS black holes~\cite{Detournay:2012pc}. The WCFT thermal retarded correlators match the scattering amplitudes in a thermal WAdS~\cite{Song:2017czq}. The holographic entanglement entropy~\cite{Song:2016gtd, Wen:2018mev, Gao:2019vcc, Apolo:2020bld, Apolo:2020qjm, Detournay:2020vrd}, the mutual information~\cite{Chen:2019xpb}, and the reflected entropy~\cite{Chen:2022fte} from WCFT are holographically constructed in AdS$_3$ or WAdS. A holographic calculation of the asymptotic growth of three point coefficient in WCFT is provided on a BTZ black hole background~\cite{Song:2019txa}. See~\cite{Castro:2015csg, Azeyanagi:2018har} for another kind of holographic realization of a WCFT in the lower spin Chern-Simons theory.

When the Kerr black hole is in or near extreme, the near horizon geometry with $SL(2, R)\times U(1)$ isometry at fixed polar angle can be viewed as a quotient of WAdS~\cite{Guica:2008mu, Bredberg:2009pv}. As mentioned above, the WAdS is conjectured to be dual to the WCFT under Dirichlet-Neumann boundary conditions~\cite{Compere:2009zj}. So it is natural be believe that other than the Kerr$/$CFT, the WCFT could also be relevant to the microstate description of the Kerr black hole. When the Kerr black hole is away from extremality, the warped symmetry is not manifest in the geometry. However, in the covariant phase space, by choosing a specific set of vector fields, the covariant charges associate to these vector fields can be shown to satisfy a warped conformal algebra with non-trivial central extensions~\cite{Aggarwal:2019iay}. This result will support the WCFT as a possible dual field theory to the Kerr black hole in the non-extreme case. Along this line, we will further investigate the scattering amplitudes of perturbations on a generic non-extreme Kerr black hole background, and explain the corresponding results in terms of WCFT thermal correlators.

This paper is organized as follows. Section \ref{sec2} is a review of the properties of the Kerr black hole as well as the corresponding hidden conformal symmetry. In section \ref{sec3}, the solutions to the wave equation of a massless scalar perturbation are discussed in the near and far regions, respectively. The retarded Green's function and the absorption probability of the scalar flux at infinity are calculated by imposing an ingoing boundary condition at the horizon of the Kerr black hole. The holographic descriptions of the retarded Green's function as well as the absorption probability are carried out in the WCFT. In section \ref{sec4}, we generalize the analysis to higher spin perturbations on the Kerr black hole background. Section \ref{sec5} is for summary and discussion.

\section{Review of Kerr spacetime and hidden conformal symmetry}\label{sec2}
In the Boyer-Lindquist coordinates, the Kerr black hole metric can be written as
\be
ds^2=-\frac{\Delta}{\rho^2}(dt-a\sin^2\theta d\phi)^2+\frac{\sin^2\theta}{\rho^2}\left((r^2+a^2)d\phi-adt\right)^2+\frac{\rho^2}{\Delta}dr^2
+\rho^2d\theta^2\,,
\ee
where
\be
\Delta=r^2-2Mr+a^2,~~~\rho^2=r^2+a^2\cos^2\theta\,,
\ee
$M$ is the black hole mass, and $a=J/M$ is the black hole angular momentum per unit mass. The coordinates' ranges are $t\in(-\infty, +\infty)$, $r\in[0, +\infty)$, $\theta\in[0, \pi]$, and $\phi\in[0, 2\pi)$. The outer $(+)$ and the inner $(-)$ horizons are located at
\be
r_{\pm}=M\pm\sqrt{M^2-a^2}\,.
\ee
Consider a massless scalar filed $\Phi$ on the Kerr black hole background. The Klein-Gordon equation is
\be
\frac{1}{\sqrt{-g}}\p_{\mu}(\sqrt{-g}g^{\mu\nu}\p_{\nu}\Phi)=0\,.
\ee
Note that both $\p_t$ and $\p_{\phi}$ are Killing vectors, the scalar field can be expanded by the following eigenmodes
\be
\Phi(t, r, \theta, \phi)=e^{-i\omega t+im\phi}S(\theta)R(r)\,,
\ee
where $\omega$ and $m$ are the frequency and angular momentum of the scalar field. Due to the $2\pi$ periodicity in coordinate $\phi$, $m$ takes integral values. With these eigenmodes, the Klein-Gordon equation splits into
\be
\left[\frac{1}{\sin\theta}\p_{\theta}(\sin\theta\p_{\theta})-\frac{m^2}{\sin^2\theta}+\omega^2a^2\cos^2\theta\right]S(\theta)=-K_{\ell}S(\theta)\,,
\ee
and
\be\label{Reqf}
\left[\frac{d}{dr}\Delta\frac{d}{dr}+\frac{(2Mr_+\omega-am)^2}{(r-r_+)(r_+-r_-)}-\frac{(2Mr_-\omega-am)^2}{(r-r_-)(r_+-r_-)}+(r^2+2M(r+2M))\omega^2\right]R(r)=K_{\ell}R(r)\,.
\ee
The above equations can be solved by Heun functions with $K_{\ell}$ be the separation constant which are the eigenvalues on a sphere.

We will consider the case in which the wave frequency is much less than the inverse of the black hole mass
\be
\omega\ll\frac{1}{M}\,.
\ee
The region $r\ll\frac{1}{\omega}$ is called the near region, and the region $r\gg M$ is called the far region. When the frequency $\omega$ is small, the near region can extend to the far region with a large overlap. The region $M\ll r\ll\frac{1}{\omega}$ is called the matching region. The hidden conformal symmetry shows up in the solution space of the wave equation for the propagating field in the near region~\cite{Castro:2010fd}.

In the near region, the angular equation with $\omega a<\omega M\ll1$ reduces to the standard Laplacian equation on a 2-sphere
\be\label{Aeq}
\left[\frac{1}{\sin\theta}\p_{\theta}(\sin\theta\p_{\theta})-\frac{m^2}{\sin^2\theta}\right]S(\theta)=-\ell(\ell-1)S(\theta)\,,~~~~\ell=-m+1,\cdots,m+1\,.
\ee
while the radial wave equation becomes
\be\label{Req}
\left[\frac{d}{dr}\Delta\frac{d}{dr}+\frac{(2Mr_+\omega-am)^2}{(r-r_+)(r_+-r_-)}-\frac{(2Mr_-\omega-am)^2}{(r-r_-)(r_+-r_-)}\right]R(r)=\ell(\ell-1)R(r)\,.
\ee
To see the structure of the wave equation more clearly, it is convenient to define the ``conformal'' coordinates~\cite{Castro:2010fd}
\bea
\omega^+&=&\sqrt{\frac{r-r_+}{r-r_-}}e^{2\pi T_R\phi}\,,\nn\\
\omega^-&=&\sqrt{\frac{r-r_+}{r-r_-}}e^{2\pi T_L\phi-\frac{t}{2M}}\,,\\
y&=&\sqrt{\frac{r_+-r_-}{r-r_-}}e^{\pi(T_L+T_R)\phi-\frac{t}{4M}}\,,\nn
\eea
where
\be
T_R=\frac{r_+-r_-}{4\pi a}\,,~~~~T_L=\frac{r_++r_-}{4\pi a}\,.
\ee
Define the local vector fields
\bea
H_1&=&i\p_+\,,\nn\\
H_0&=&i(\omega^+\p_++\frac{1}{2}y\p_y)\,,\\
H_{-1}&=&i(\omega^{+2}\p_++\omega^+y\p_y-y^2\p_-)\,,\nn
\eea
and
\bea
\bar{H}_1&=&i\p_-\,,\nn\\
\bar{H}_0&=&i(\omega^-\p_-+\frac{1}{2}y\p_y)\,,\\
\bar{H}_{-1}&=&i(\omega^{-2}\p_-+\omega^-y\p_y-y^2\p_+)\,.\nn
\eea
The above vector fields form the $SL(2, R)$ Lie algrbra
\be
[H_0, H_{\pm1}]=\mp iH_{\pm1}\,,~~~~[H_{-1}, H_1]=-2iH_0\,.
\ee
A similar relation holds for $(\bar{H}_0, \bar{H}_{\pm1})$. The corresponding $SL(2, R)$ quadratic Casimir operator is
\be\label{cc}
\mathcal{H}^2=-H_0^2+\frac{1}{2}(H_1H_{-1}+H_{-1}H_1)\,.
\ee
One can check that
\be\label{ce}
\mathcal{H}^2=\p_r\Delta\p_r-\frac{(2Mr_+\p_t+a\p_{\phi})^2}{(r-r_+)(r_+-r_-)}+\frac{(2Mr_-\p_t+a\p_{\phi})^2}{(r-r_-)(r_+-r_-)}\,.
\ee
So in the near region, the scalar field with fixed $\theta$ coordinate satisfies the following equation
\be
\mathcal{H}^2\Phi=\ell(\ell-1)\Phi\,,
\ee
which indicate that the $\Phi$ is within the global conformal representation of the $SL(2, R)$ with conformal weight $\ell$. This equation have both right and left copy from $(H_0, H_{\pm1})$ and $(\bar{H}_0, \bar{H}_{\pm1})$. The $SL(2, R)\times SL(2, R)$ symmetry is spontaneously broken by the angular identification of the $\phi$ coordinate, this requires the corresponding 2D CFT at temperature
\be
T_R=\frac{r_+-r_-}{4\pi a}\,,~~~~T_L=\frac{r_++r_-}{4\pi a}\,.
\ee
The above statement is supported by the black hole entropy, scattering amplitudes~\cite{Castro:2010fd}, the covariant phase space formalism~\cite{Haco:2018ske} and the gravitational radiations~\cite{Nian:2023dng}.

\section{Scalar radiations in Kerr spacetime}\label{sec3}
\subsection{Near region}
As mentioned in~\cite{Aggarwal:2019iay}, the potential term in the near region radial wave equation \eqref{Reqf} can be approximately written with an additional $4M^2\omega^2$ term due to the $r\omega$ terms have negligible contributions in the frequency space. Taking this into account, the near region radial wave equation becomes
\be\label{Reqlp}
\left[\frac{d}{dr}\Delta\frac{d}{dr}+\frac{(2Mr_+\omega-am)^2}{(r-r_+)(r_+-r_-)}-\frac{(2Mr_-\omega-am)^2}{(r-r_-)(r_+-r_-)}\right]R(r)=\ell'(\ell'-1)R(r)\,,
\ee
where
\be\label{ellp}
\ell'=\frac{1}{2}+\frac{1}{2}\sqrt{(2\ell-1)^2-16M^2\omega^2}\,.
\ee
The $SL(2, R)$ Casimir equation still holds
\be
\mathcal{H}^2\Phi=\ell'(\ell'-1)\Phi,,
\ee
but now with a frequency dependent conformal weight $\ell'$. We will assign the frequency $\omega$ being proportional to the $U(1)$ charge $Q$ of a WCFT. The charge dependence of the conformal weight in holographic WCFT also happens in some other contexts. For example, in the studying of the holographic interpretation of bulk scalar Green's functions on a thermal WAdS background in terms of WCFT data~\cite{Song:2017czq}, the conformal weight of the scalar field depends on the charge of the dual WCFT. In the present paper, we are going to phenomenologically determine the scattering amplitudes in the Kerr spacetime in terms of WCFT correlators. We will keep using $\ell'$ instead of $\ell$ in the following discussion.

In the near region, the radial wave equation \eqref{Reqlp} can be solved by the hypergeometric functions. Imposing an ingoing boundary condition at the horizon, the solution can be written as
\bea\label{Rn}
R(r)&=&\left(\frac{r-r_+}{r-r_-}\right)^{-i\frac{2Mr_+\omega-am}{r_+-r_-}}(r-r_-)^{-\ell'}\nn\\
&&\times_2F_1\left(\ell'-i\frac{4M^2\omega-2am}{r_+-r_-}, \ell'-i2M\omega; 1-i\frac{4Mr_+\omega-2am}{r_+-r_-}; \frac{r-r_+}{r-r_-}\right)\,,
\eea
where $_2F_1(a, b; c; z)$ is the hypergeometric function. When the frequency $\omega$ is very small, the near region $r\ll\frac{1}{\omega}$ extends to the far region where the asymptotic behavior of the scalar field can be analysed. The near region radial wave solution in the far region behaves as
\be\label{Rni}
R(r\gg M)\sim A_{near}r^{-\ell'}+B_{near}(r_+-r_-)^{-2\ell'+1}r^{\ell'-1}\,,
\ee
where
\bea
A_{near}&=&\frac{\Gamma\left(1-i\frac{4Mr_+\omega-2am}{r_+-r_-}\right)\Gamma(-2\ell'+1)}{\Gamma(1-\ell'-i2M\omega)\Gamma\left(1-\ell'-i\frac{4M^2\omega-2am}{r_+-r_-}\right)}\,,\label{An}\\
B_{near}&=&\frac{\Gamma\left(1-i\frac{4Mr_+\omega-2am}{r_+-r_-}\right)\Gamma(2\ell'-1)}{\Gamma(\ell'-i2M\omega)\Gamma\left(\ell'-i\frac{4M^2\omega-2am}{r_+-r_-}\right)}\,.\label{Bn}
\eea
The retarded Green's function for the scalar field $\Phi$ in the frequency space can be read off from the asymptotic expansion
\be\label{GRKerr}
G_R^{bulk}\sim\frac{A_{near}}{B_{near}}=\frac{\Gamma(-2\ell'+1)}{\Gamma(2\ell'-1)}\frac{\Gamma(\ell'-i2M\omega)}{\Gamma(1-\ell'-i2M\omega)}\frac{\Gamma\left(\ell'-i\frac{4M^2\omega-2am}{r_+-r_-}\right)}{\Gamma\left(1-\ell'-i\frac{4M^2\omega-2am}{r_+-r_-}\right)}\,,
\ee
here $\sim$ stands for equaling up to a factor independent of $\omega$ and $m$.

\subsection{Far region}
In the far region, the redial wave equation reduces to
\be
\frac{d}{dr}\left(r^2\frac{dR(r)}{dr}\right)+r^2\omega^2R(r)=\ell'(\ell'-1)R(r)\,.
\ee
This is the spherical Bessel's equation. There are two linearly independent solutions~\cite{Nian:2023dng}
\be\label{Rf}
R(r)=A_{far}\frac{1}{\sqrt{\omega r}}\mathrm{J}_{-\ell'+\frac{1}{2}}(\omega r)+B_{far}\frac{1}{\sqrt{\omega r}}\mathrm{J}_{\ell'-\frac{1}{2}}(\omega r)\,,
\ee
where $A_{far}$ and $B_{far}$ are constants. These solutions have the following asymptotic behaviors
\bea
\frac{1}{\sqrt{\omega r}}\mathrm{J}_{-\ell'+\frac{1}{2}}(\omega r)&\to&\omega^{-\ell'}r^{-\ell'}\frac{2^{\ell'-\frac{1}{2}}}{\Gamma\left(-\ell'+\frac{3}{2}\right)}\,,~~~~r\to0\,,\label{Rfa0}\\
&\to&\sqrt{\frac{2}{\pi}}\frac{1}{\omega r}\sin(\omega r+\pi\ell'/2)\,,~~~~r\to\infty\,,\label{Rfai}\\
\frac{1}{\sqrt{\omega r}}\mathrm{J}_{\ell'-\frac{1}{2}}(\omega r)&\to&\omega^{\ell'-1}r^{\ell'-1}\frac{2^{-\ell'+\frac{1}{2}}}{\Gamma\left(\ell'+\frac{1}{2}\right)}\,,~~~~r\to0\,,\label{Rfb0}\\
&\to&\sqrt{\frac{2}{\pi}}\frac{1}{\omega r}\cos(\omega r-\pi\ell'/2)\,,~~~~r\to\infty\,.\label{Rfbi}
\eea

\subsection{Near-far matching}
At the outer boundary of the far region, each of these two solutions has outgoing and ingoing modes. One can separate the pure outgoing and pure ingoing part in the solution with the following recombination
\bea
R(r)&=&Z_{out}\left(ie^{i\pi\ell'}\frac{1}{\sqrt{\omega r}}\mathrm{J}_{-\ell'+\frac{1}{2}}(\omega r)+\frac{1}{\sqrt{\omega r}}\mathrm{J}_{\ell'-\frac{1}{2}}(\omega r)\right)\nn\\
&&+Z_{in}\left(-ie^{-i\pi\ell'}\frac{1}{\sqrt{\omega r}}\mathrm{J}_{-\ell'+\frac{1}{2}}(\omega r)+\frac{1}{\sqrt{\omega r}}\mathrm{J}_{\ell'-\frac{1}{2}}(\omega r)\right)\,,
\eea
where
\bea
Z_{out}&=&\frac{1}{2\cos(\pi\ell')}(-iA_{far}+e^{-i\pi\ell'}B_{far})\,,\\
Z_{in}&=&\frac{1}{2\cos(\pi\ell')}(iA_{far}+e^{i\pi\ell'}B_{far})\,,
\eea
are the amplitudes of the pure outgoing and ingoing part, respectively. With the asymptotic behaviors given in \eqref{Rfai} and \eqref{Rfbi}, one can show that the scalar fields at the outer boundary of the far region take the forms
\bea
\Phi_{out}(r\to\infty)&=&\sum_{\ell, m}e^{-i\omega t+im\phi}S_{\ell}(\theta)Z_{out}\left(\sqrt{\frac{2}{\pi}}\frac{e^{i\omega r+i\pi\ell'/2}}{\omega r}\cos(\pi\ell')\right)\,,\\
\Phi_{in}(r\to\infty)&=&\sum_{\ell, m}e^{-i\omega t+im\phi}S_{\ell}(\theta)Z_{in}\left(\sqrt{\frac{2}{\pi}}\frac{e^{-i\omega r-i\pi\ell'/2}}{\omega r}\cos(\pi\ell')\right)\,.
\eea
The Klein-Gordon particle number flux is defined as
\be
\mathcal{F}=\int\sqrt{-g}J^rd\theta d\phi\,,~~~~\text{with}~J^{\mu}=\frac{i}{8\pi}(\Phi^*\nabla^{\mu}\Phi-\Phi\nabla^{\mu}\Phi^*)\,.
\ee
At infinity, the fluxes of outgoing and ingoing particle number behave like
\be
\mathcal{F}_{out}\propto|Z_{out}|^2\,,~~~~\mathcal{F}_{in}\propto|Z_{in}|^2\,,
\ee
with same proportional constants. The absorption probability, which is the ratio of the net incoming flux to the total incoming flux, can be expressed as
\be\label{absKerrex}
\sigma_{abs}^{bulk}=\frac{\mathcal{F}_{in}-\mathcal{F}_{out}}{\mathcal{F}_{in}}=1-\frac{|Z_{out}|^2}{|Z_{in}|^2}
=\frac{2i\cos(\pi\ell')(A_{far}B_{far}^*-B_{far}A_{far}^*)}{|iA_{far}+e^{i\pi\ell'}B_{far}|^2}\,.
\ee
The radial solutions at far and near regions should match with each other at the matching region. The far region $R(r)$ at $r\to0$ should have the same coefficients with the near region $R(r)$ at $r\gg M$. By comparing with \eqref{Rf}, \eqref{Rfa0}, \eqref{Rfb0} and \eqref{Rni}, these coefficients are related through
\bea
A_{far}&=&2^{-\ell'+\frac{1}{2}}\Gamma\left(-\ell'+\frac{3}{2}\right)\omega^{\ell'}A_{near}\,,\\
B_{far}&=&2^{\ell'-\frac{1}{2}}\Gamma\left(\ell'+\frac{1}{2}\right)(r_+-r_-)^{-2\ell'+1}\omega^{-\ell'+1}B_{near}\,.
\eea
Using \eqref{An} and \eqref{Bn}, and the above relations, we find
\be
2i\cos(\pi\ell')(A_{far}B_{far}^*-B_{far}A_{far}^*)=\pi\omega(r_+-r_-)^{-2\ell'}(4Mr_+\omega-2am)\,.
\ee
Note that we assume $\omega M\ll1$, and $r_+-r_-$ is at order of $M$, the coefficient $B_{far}$ will be much larger than $A_{far}$ when $\mathrm{Re}(\ell')>1/2$. So the absorption probability in this case can be written as
\bea\label{absKerr}
\sigma_{abs}^{bulk}&=&\frac{2i\cos(\pi\ell')(A_{far}B_{far}^*-B_{far}A_{far}^*)}{|B_{far}|^2}\nn\\
&=&\frac{(\omega/2)^{2\ell'-1}(r_+-r_-)^{2\ell'-1}}{\Gamma\left(\ell'+\frac{1}{2}\right)^2\Gamma(2\ell'-1)^2}\sinh\left(\pi\frac{4Mr_+\omega-2am}{r_+-r_-}\right)\nn\\
&\times&|\Gamma(\ell'-i2M\omega)|^2\left|\Gamma\left(\ell'-i\frac{4M^2\omega-2am}{r_+-r_-}\right)\right|^2\,.
\eea
It is clear that when the modes parameters $\omega$ and $m$ satisfying
\be
0<\omega<m\frac{a}{2Mr_+}\,,
\ee
the absorption probability is negative, which means that the outgoing flux is more energetic than the incoming one. This corresponds to the superradiant scattering case~\cite{Zeldovich:1971, Zeldovich:1972}.

In the next section, we will calculate the thermal retarded correlator and the absorption cross section of the WCFT, and show their matchings with the bulk quantities \eqref{GRKerr} and \eqref{absKerr}, respectively. The holographic dual calculation of the absorption probability only applies for $\mathrm{Re}(\ell')>1/2$. When this condition is not satisfied, i.e. $\mathrm{Re}(\ell')=1/2$, both $A_{far}$ and $B_{far}$ are relevant in the denominator of \eqref{absKerrex}. The dependence of the bulk absorption probability on the small frequency becomes highly oscillating.

\subsection{Dual calculations in WCFT}
For a WCFT with coordinates $(x, y)$,  the generic warped conformal symmetry transformation can be written as
\be
x'=f(x)\,,~~~~y'=y+g(x)\,,
\ee
where $f(x)$ and $g(x)$ are two arbitrary functions. First, consider the zero temperature case in which the WCFT is defined on a plane. Infinitesimally, the warped conformal symmetries are generated by a set of vector fields
\be
L_n=-x^{n+1}\p_x\,,~~~~P_n=ix^n\p_y\,.
\ee
These vector fields form the Virasoro-Kac-Moody algebra
\begin{align}
[L_n, L_m]&=(n-m)L_{n+m}\,,\nn\\
[L_n, P_m]&=-mP_{n+m}\,,\nn\\
[P_n, P_m]&=0\,.
\end{align}
The global symmetries are generated by $(L_{\pm1}, L_0, P_0)$ which form a $sl(2, R)\times u(1)$ subalgebra. At charge level, there are infinitely many conserved charges associated with translations along $x$ and $y$, with corresponding Neother currents $T(x)$ and $P(x)$, respectively
\be
\mathcal{L}_n=-\frac{i}{2\pi}\int dxx^{n+1}T(x)\,,~~~~\mathcal{P}_n=-\frac{1}{2\pi}\int dxx^nP(x)\,.
\ee
The commutation relations for the charges form a warped conformal algebra consists of one Virasoro algebra and one $U(1)$ Kac-Moody algebra with central extension~\cite{Detournay:2012pc}
\begin{align}
[\mathcal{L}_n, \mathcal{L}_m]&=(n-m)\mathcal{L}_{n+m}+\frac{c}{12}n(n^2-1)\delta_{n, -m}\,,\nn\\
[\mathcal{L}_n, \mathcal{P}_m]&=-m\mathcal{P}_{n+m}\,,\nn\\
[\mathcal{P}_n, \mathcal{P}_m]&=k\frac{n}{2}\delta_{n, -m}\,,
\end{align}
where $c$ is the central charge and $k$ is the Kac-Moody level. Similar to the CFT story, one can define the primary operators of the WCFT through the following transformation law~\cite{Song:2017czq}
\be
\mathcal{O}'(x', y')=\left(\frac{\p x'}{\p x}\right)^{-\delta}\mathcal{O}(x, y)\,,
\ee
where $\delta$ is the scaling dimension of the primary operator $\mathcal{O}(x, y)$. The infinitesimal version can be expressed as
\begin{align}
[\mathcal{L}_n, \mathcal{O}(x, y)]&=[x^{n+1}\p_x+(n+1)x^n\delta]\mathcal{O}(x, y)\,,\\
[\mathcal{P}_n, \mathcal{O}(x, y)]&=ix^n\p_y\mathcal{O}(x, y)=-x^nQ\mathcal{O}(x, y)\,,
\end{align}
where the last equation is because we are choosing a basis with definite $\mathcal{P}_0$ charge $Q$. The quadratic Casimir operator acting on $\mathcal{O}(x, y)$ as an eigenoperator with eigenvalue $\delta(\delta-1)$
\be
L^2\mathcal{O}=\left(L_0-\frac{1}{2}(L_1L_{-1}+L_{-1}L_1)\right)\mathcal{O}=\delta(\delta-1)\mathcal{O}\,.
\ee

The finite temperature case can be obtained by applying a warped conformal transformation~\cite{Song:2017czq}
\be
x=e^{\frac{2\pi X}{\beta}}\,,~~~~y=Y+\bar{q}X\,.
\ee
$(X, Y)$ stands for the finite temperature coordinates with the following thermal identification
\be
(X, Y)\sim(X+i\beta, Y-i\bar{\beta})\,,
\ee
where $\beta$ is the inverse temperature along $X$ and $\bar{\beta}=\bar{q}\beta+\alpha$ is the inverse temperature along $Y$. $\bar{q}$ is the spectral flow parameter. In terms of $(X, Y)$, the warped conformal generators can be written as
\be\label{LP}
L_n=-\frac{\beta}{2\pi}e^{\frac{2\pi nX}{\beta}}(\p_X-\bar{q}\p_Y)\,,~~~~P_n=ie^{\frac{2\pi nX}{\beta}}\p_Y\,.
\ee

We relate the finite temperature spectral flowed WCFT coordinates $(X, Y)$ to the Kerr coordinates $(t, \phi)$ in the following way
\be\label{tX}
X=2\pi T_R\phi\,,~~~~Y=\frac{1}{2M}t-2\pi(T_L+\bar{q}T_R)\phi\,.
\ee
The angular identification $\phi\sim\phi+2\pi$ can be expressed as
\be
(X, Y)\sim\left(X+4\pi^2T_R, Y-4\pi^2(T_L+\bar{q}T_R)\right)\,.
\ee
Further more, due to the facts that the Kerr black hole has temperature and the corresponding coordinates have the following thermal identification
\be
(t, \phi)\sim(t+iT_H^{-1}, \phi+iT_H^{-1}\Omega_H)\,,
\ee
where $T_H$ and $\Omega_H$ are the Hawking temperature and the angular velocity of the outer horizon
\be
T_H=\frac{r_+-r_-}{8\pi Mr_+}\,,~~~~\Omega_H=\frac{a}{2Mr_+}\,.
\ee
Then the WCFT coordinates inherit the following thermal identification
\be
(X, Y)\sim\left(X+2\pi i, Y-2\pi i(\bar{q}-1)\right)\,,
\ee
with $\beta=2\pi$ and $\bar{\beta}=2\pi(\bar{q}-1)$. So the finite temperature WCFT is defined on a torus with both spatial and thermal identifications
\be
(X, Y)\sim\left(X+4\pi^2T_R, Y-4\pi^2(T_L+\bar{q}T_R)\right)\sim\left(X+2\pi i, Y-2\pi i(\bar{q}-1)\right)\,.
\ee

At low frequency, the near region $r\ll\frac{1}{\omega}$ extends to very far at $r\gg M$. The WCFT lives on this fixed large $r$ surface. In this place, the conformal vector fields $(H_{\pm1}, H_0, \bar{H}_0)$ which form $sl(2, R)\times u(1)$ subalgebra can be identified as the global part of the warped conformal generators $(L_{\pm1}, L_0, P_0)$ in \eqref{LP} by using \eqref{tX}, through
\be\label{LH}
L_{\pm1}=iH_{\mp1}\,,~~~~L_0=iH_0\,,~~~~P_0=-\bar{H}_0\,.
\ee
With the above relation, the quadratic Casimir operator $L^2$ of the $SL(2, R)$ in the warped conformal algebra
\be
L^2=L_0^2-\frac{1}{2}(L_1L_{-1}+L_{-1}L_1)=\mathcal{H}^2\,,
\ee
is relate to the conformal Casimir operator $\mathcal{H}^2$ in the scalar wave equation \eqref{ce}. Thus we can relate the scaling dimension $\delta$ of the WCFT primary operator to the $SL(2, R)$ conformal weight of the scalar field $\Phi$
\be\label{dl}
\delta=\ell'=\frac{1}{2}+\frac{1}{2}\sqrt{(2\ell-1)^2-16M^2\omega^2}\,.
\ee

The action of the deformed WCFT can be written as
\be
S=S_{WCFT}+\sum_{\delta}\int dXdYJ_{\delta, Q}(X, Y)\mathcal{O}_{\delta, Q}(X, Y)\,,
\ee
where $S_{WCFT}$ is the original WCFT action, and $\mathcal{O}_{\delta, Q}(X, Y)$ are the WCFT operator with scaling dimension $\delta$ and fix $\mathcal{P}_0$ charge $Q$. The source term, which comes from the unnormalized part of the asymptotic scalar field in the Kerr spacetime, can be expanded in terms of the periodic $\phi$ modes
\be
J_{\delta, Q}(X, Y)=\sum_mJ_{\delta m}e^{-i\Omega X+iQY}\,,
\ee
where $J_{\delta m}\propto B_{near}$, and
\be\label{OQ}
\Omega=\frac{4M^2\omega-2am}{r_+-r_-}+2M\omega\bar{q}\,,~~~~Q=-2M\omega\,.
\ee
Here we view $Q$ as fixed constant and $\Omega$ is the frequency along $X$.

Inspired by the CFT's result, the Fermi's Golden rule gives the transition rate out of the thermal state~\cite{Maldacena:1997ih, Gubser:1997cm}
\be\label{tr}
\mathcal{R}=2\pi\sum_{\delta m}|J_{\delta m}|^2\int dXdYe^{i\Omega X-iQY}G(X, Y)\,,
\ee
where $G(X, Y)$ is the finite temperature two point function of the WCFT~\cite{Song:2017czq}
\be\label{2pf}
G(X, Y)=\langle\mathcal{O}_{\delta, Q}(X, Y)\mathcal{O}_{\delta, -Q}(0, 0)\rangle\sim\mathcal{C}_{\mathcal{O}}(-1)^{\delta}e^{iQ(Y+\bar{q}X)}\left(\frac{\beta}{\pi}\sinh\frac{\pi X}{\beta}\right)^{-2\delta}\,,
\ee
with $\beta=2\pi$ and $\mathcal{C}_{\mathcal{O}}$ being the normalization constant. From the transition rate \eqref{tr}, one can write down the thermal retarded correlator and the absorption cross section in the following form~\cite{Bredberg:2009pv}
\bea
G_R&\sim&\int dXdYe^{i\Omega X-iQY}G(X-i\epsilon, Y-i\epsilon)\,,\label{FGG}\\
\sigma_{abs}&\sim&\int dXdYe^{i\Omega X-iQY}[G(X-i\epsilon, Y-i\epsilon)-G(X+i\epsilon, Y+i\epsilon)]\,,\label{FGs}
\eea
where $-i\epsilon$ and $+i\epsilon$ correspond to absorption and emission. Using \eqref{2pf} in \eqref{FGG} and further notice that the Euclidean frequency at finite temperature takes discrete values of the Matsubara frequencies, the thermal retarded correlator then takes the from~\cite{Song:2017czq}
\be\label{GRWCFT}
G_R\sim(2\pi)^{1-2\delta}\frac{e^{\pi Q\bar{q}}}{\sin\left(\pi(\delta+iQ\bar{q})\right)}\frac{\Gamma\left(\delta-i(\Omega+Q\bar{q})\right)}{\Gamma\left(1-\delta-i(\Omega+Q\bar{q})\right)}\,,
\ee
where we have set $\beta=2\pi$. Invoking the relations \eqref{dl} and \eqref{OQ}, one can show that the retarded Green's function for the scalar filed in the Kerr spacetime \eqref{GRKerr} matches the WCFT thermal retarded correlator \eqref{GRWCFT} in the $\Omega$ dependence.

Substituting \eqref{2pf} into \eqref{FGs} and integrating over the subtraction, the absorption cross section takes the form
\bea
\sigma_{abs}&\sim&\frac{1}{\Gamma(2\delta)}\sinh(\pi(\Omega+Q\bar{q}))|\Gamma(\delta-i(\Omega+Q\bar{q}))|^2\nn\\
&\sim&\frac{1}{\Gamma(2\delta)}\sinh(\pi(\Omega+Q\bar{q}-Q))|\Gamma(\delta-i(\Omega+Q\bar{q}))|^2+\mathcal{O}(Q)\,.\label{absWCFT}
\eea
Here we expand the $\sinh$ function in terms of small charge up to leading order since the charge $Q$ is proportional to the low frequency $\omega$ as in \eqref{OQ}. Given the relations \eqref{dl} and \eqref{OQ}, the absorption probability of the scalar flux in the Kerr spacetime at low frequency \eqref{absKerr} matches the WCFT absorption cross section \eqref{absWCFT} at small charge in the $\Omega$ dependence. So the WCFT calculation covers the leading $\Omega$ dependence in the bulk absorption probability.

The WCFT calculations reproduce the right-moving $\Omega+Q\bar{q}$ dependent parts of the bulk retarded Green's function and absorption probability. The left-moving parts which only depend on the charge $Q$ are missing. This is because the WCFT contains only one Virasoro algebra at symmetry level. However, this is not an issue since the charge $Q$ is viewed as a fixed constant in the WCFT. The $Q$ dependent functions can be normalized in the operator's definition. One need another Virasoro sector in the two point function to recover the $Q$ dependent left-moving parts.

At symmetry level, the non-extreme Kerr black hole possesses two copies of Virasoro algebra~\cite{Haco:2018ske} in the near horizon region as well as Virasoro and Kac-Moody algebra~\cite{Aggarwal:2019iay}. Solely look at bulk Kerr spacetime geometry, one can not tell the dual field theory being CFT or WCFT. It depends on what kinds differmorphism or boundary conditions near the horizon is included. At perturbation level, the bulk quantities also allow both CFT and WCFT descriptions. It depends on the frequency $\omega$ and angular momentum $m$ of the scalar field. If both are variables, the hidden conformal symmetry ensures that they have CFT description. If $\omega$ is held as a fixed constant, then the allowed hidden conformal symmetry generators reduce to $(H_{\pm1}, H_0, \bar{H}_0)$. This is because the eigenvalue of $\bar{H}_0$ acting on the scalar field is proportional to the frequency $\omega$. However, the $\bar{H}_{\pm1}$ operating will change the eigenvalue since they have non-trivial commutation relation to $\bar{H}_0$. This is not expected from a WCFT description since the frequency is related to the fixed $U(1)$ charge $Q$ through \eqref{OQ}. Furthermore, the conformal generators $(H_{\pm1}, H_0, \bar{H}_0)$ can be identified as the global part of the warped conformal generators through \eqref{LH} and the local symmetry algebra of a WCFT can be reproduced in the near horizon region of a generic Kerr black hole~\cite{Aggarwal:2019iay}. Therefore, start from the bulk analysis of perturbations on a Kerr background, if one fix the frequency, the result is expected to be recovered form a WCFT calculation. We calculate the retarded Green's function and absorption probability in the Kerr spacetime, and found that the fixed frequency result indeed match the WCFT calculation.

\section{Higher spin radiations in Kerr spacetime}\label{sec4}
In this section, we generalize the previous discussion about the scalar field to the higher spin field case, for instance the gravitational perturbation on the Kerr background. To explore the higher spin case, it is convenient to apply the Newman-Penrose formalism, by introducing the NP null tetrad of the Kerr metric in components $(t, r, \theta, \phi)$
\bea
\ell^{\mu}&=&\left(\frac{r^2+a^2}{\Delta}, 1, 0, \frac{a}{\Delta}\right)\,,\nn\\
n^{\mu}&=&\frac{1}{2(r^2+a^2\cos^2\theta)}\left(r^2+a^2, -\Delta, 0, a\right)\,,\\
m^{\mu}&=&\frac{1}{\sqrt{2}(r+ia\cos\theta)}\left(ia\sin\theta, 0, 1, \frac{i}{\sin\theta}\right)\,.\nn
\eea
The perturbations of the following gauge invariant Weyl scalars describe the gravitational radiations
\bea
\Psi_{s=-2}&=&\Psi_4^{(1)}=C_{\mu\nu\rho\sigma}n^{\mu}m^{*\nu}n^{\rho}m^{*\sigma}\,,\\
\Psi_{s=2}&=&\Psi_0^{(1)}=C_{\mu\nu\rho\sigma}\ell^{\mu}m^{\nu}\ell^{\rho}m^{\sigma}\,,
\eea
where $\Psi_{s=-2}$ is for outgoing radiation and $\Psi_{s=2}$ is for ingoing. These perturbations satisfy the following master equation on the Kerr background~\cite{Teukolsky:1972my, Teukolsky:1973ha}
\bea
&&\left(\frac{(r^2+a^2)^2}{\Delta}-a^2\sin^2\theta\right)\frac{\p^2\Psi_s}{\p t^2}+\frac{4Mar}{\Delta}\frac{\p^2\Psi_s}{\p t\p\phi}+\left(\frac{a^2}{\Delta}-\frac{1}{\sin\theta}\right)\frac{\p^2\Psi_s}{\p\phi^2}\nn\\
&&-\Delta^{-s}\frac{\p}{\p r}\left(\Delta^{s+1}\frac{\p\Psi_s}{\p r}\right)-\frac{1}{\sin\theta}\frac{\p}{\p\theta}\left(\sin\theta\frac{\p\Psi_s}{\p\theta}\right)-2s\left(\frac{a(r-M)}{\Delta}+\frac{i\cos\theta}{\sin^2\theta}\right)\frac{\p\Psi_s}{\p\phi}\nn\\
&&-2s\left(\frac{M(r^2-a^2)}{\Delta}-r-ia\cos\theta\right)\frac{\p\Psi_s}{\p t}+(s^2\cot^2\theta-s)\Psi_s=0\,.\label{meq}
\eea
This equation is valid for a generic spin-weight parameter $s$ and we have set the source term vanishing on the right hand side. The master equation \eqref{meq} is separable by writing
\be
\Psi_s=e^{-i\omega t+im\phi}S^s(\theta)R^s(r)\,,
\ee
which leads to the following angular and radial equations
\bea
&&\frac{1}{\sin\theta}\frac{d}{d\theta}\left(\sin\theta\frac{dS^s(\theta)}{d\theta}\right)-\left(\frac{m^2}{\sin^2\theta}+\frac{2ms\cos\theta}{\sin^2\theta}+s^2\cot^2\theta-s\right)S^s(\theta)\nn\\
&&-2\omega as\cos\theta S^s(\theta)+\omega^2a^2\cos^2\theta S^s(\theta)=-K_{\ell}^sS^s(\theta)\,,
\eea
\bea
&&\Delta^{-s}\frac{d}{dr}\left(\Delta^{s+1}\frac{dR^s(r)}{dr}\right)+\left(\frac{(2Mr_+\omega-am)^2}{(r-r_+)(r_+-r_-)}-\frac{(2Mr_-\omega-am)^2}{(r-r_-)(r_+-r_-)}\right)R^s(r)\nn\\
&&-\left(is\frac{2Mr_+\omega-am}{r-r_+}+is\frac{2Mr_-\omega-am}{r-r_-}\right)R^s(r)+2is\omega(r-M)R^s(r)\nn\\
&&+(r^2+2M(r+2M))\omega^2R^s(r)=K_{\ell}^sR^s(r)\,,
\eea
with $K^s_{\ell}$ being the separation constant. The above two equations are actually confluent Heun equations both with two regular singular points and an irregular singular point. The exact solutions can be analysed via mapping these equations to the level 2 null-state equation for primary operators of Virasoro algebra~\cite{Bonelli:2021uvf}. In this paper, we will focus on the low frequency limit case.

\subsection{Near region}
In the near region and low frequency limit, the angular equation becomes
\be
\frac{1}{\sin\theta}\frac{d}{d\theta}\left(\sin\theta\frac{dS^s(\theta)}{d\theta}\right)-\left(\frac{m^2}{\sin^2\theta}+\frac{2ms\cos\theta}{\sin^2\theta}+s^2\cot^2\theta-s\right)S^s(\theta)=-K^s_{\ell}S^s(\theta)\,,
\ee
which is the angular part of spin weighted spherical harmonic equation with $K^s_{\ell}=(\ell+s)(\ell-s-1)$~\cite{Goldberg:1967}. And the radial equation reduces to
\bea
&&\Delta^{-s}\frac{d}{dr}\left(\Delta^{s+1}\frac{dR^s(r)}{dr}\right)+\left(\frac{(2Mr_+\omega-am)^2}{(r-r_+)(r_+-r_-)}-\frac{(2Mr_-\omega-am)^2}{(r-r_-)(r_+-r_-)}\right)R^s(r)\nn\\
&&-\left(is\frac{2Mr_+\omega-am}{r-r_+}+is\frac{2Mr_-\omega-am}{r-r_-}\right)R^s(r)=K^s_{\ell'}R^s(r)\,.
\eea
Here we preserve the additional $4M^2\omega^2$ term in the potential and keep using the $\omega$ deformed $\ell'$ as in \eqref{ellp} in stead of $\ell$ in the radial equation. The solution to the radial equation in the near region can be found by the Green function method~\cite{Nian:2023dng}. Imposing an ingoing boundary condition at the horizon, the radial solution can be expressed as
\bea\label{Rns}
&&R^s(r)=\left(\frac{r-r_+}{r-r_-}\right)^{-i\frac{2Mr_+\omega-am}{r_+-r_-}}(r-r_-)^{-\ell'-s}\nn\\
&&\times_2F_1\left(\ell'-i\frac{4M^2\omega-2am}{r_+-r_-}, \ell'-s-i2M\omega; 1-s-i\frac{4Mr_+\omega-2am}{r_+-r_-}; \frac{r-r_+}{r-r_-}\right)\,.
\eea
When the $\omega$ is very small, the near region extends to the far region. At the outer boundary of the near region, the above solution behaves as
\be\label{Rnsi}
R^s(r\gg M)\sim A^s_{near}r^{-\ell'-s}+B^s_{near}(r_+-r_-)^{-2\ell'+1}r^{\ell'-s-1}\,,
\ee
where
\bea
A^s_{near}&=&\frac{\Gamma\left(1-s-i\frac{4Mr_+\omega-2am}{r_+-r_-}\right)\Gamma(-2\ell'+1)}{\Gamma(1-\ell'-s-i2M\omega)\Gamma\left(1-\ell'-i\frac{4M^2\omega-2am}{r_+-r_-}\right)}\,,\label{Ans}\\
B^s_{near}&=&\frac{\Gamma\left(1-s-i\frac{4Mr_+\omega-2am}{r_+-r_-}\right)\Gamma(2\ell'-1)}{\Gamma(\ell'-s-i2M\omega)\Gamma\left(\ell'-i\frac{4M^2\omega-2am}{r_+-r_-}\right)}\,.\label{Bns}
\eea
The retarded Green's function for the spin-s field $\Psi_s$ in the frequency space then proportional to the ratio
\be\label{GRsKerr}
G_R^{s, bulk}\sim\frac{A^s_{near}}{B^s_{near}}=\frac{\Gamma(-2\ell'+1)}{\Gamma(2\ell'-1)}\frac{\Gamma(\ell'-s-i2M\omega)}{\Gamma(1-\ell'-s-i2M\omega)}\frac{\Gamma\left(\ell'-i\frac{4M^2\omega-2am}{r_+-r_-}\right)}{\Gamma\left(1-\ell'-i\frac{4M^2\omega-2am}{r_+-r_-}\right)}\,.
\ee

\subsection{Far region}
In the far region, the radial equation reduces to
\be
r^{-2s}\frac{d}{dr}\left(r^{2s+2}\frac{dR^s(r)}{dr}\right)+(r^2\omega^2+2isr\omega)R^s(r)=(\ell'+s)(\ell'-s-1)R^s(r)\,.
\ee
The solutions of the above equation are the confluent hypergeometric functions~\cite{Nian:2023dng}. The generic solution can be expressed as
\be
R^s(r)=A^s_{far}R^s_1+B^s_{far}R^s_2\,,
\ee
where $A^s_{far}$ and $B^s_{far}$ are constants, and
\bea
R^s_1&=&r^{-\ell'-s}e^{-i\omega r}~_1F_1(-\ell'-s+1, -2\ell'+2; 2i\omega r)\,,\\
R^s_2&=&r^{\ell'-s-1}e^{-i\omega r}~_1F_1(\ell'-s, 2\ell'; 2i\omega r)\,.
\eea
The asymptotic behaviors of $R^s_1$ and $R^s_2$ are
\bea
R^s_1&\to&r^{-\ell'-s}\,,~~~~r\to0\,,\\
&\to&\alpha^s_{in}r^{-1}e^{-i\omega r}+\alpha^s_{out}r^{-2s-1}e^{i\omega r}\,,~~~~r\to\infty\,,\\
R^s_2&\to&r^{\ell'-s-1}\,,~~~~r\to0\,,\\
&\to&\beta^s_{in}r^{-1}e^{-i\omega r}+\beta^s_{out}r^{-2s-1}e^{i\omega r}\,,~~~~r\to\infty\,,
\eea
where
\bea
\alpha^s_{in}&=&\frac{(2)^{\ell'+s-1}(-i\omega)^{\ell'+s-1}\Gamma(2-2\ell')}{\Gamma(-\ell'+s+1)}\,,~~~~\alpha^s_{out}=\frac{(2i\omega)^{\ell'-s-1}\Gamma(2-2\ell')}{\Gamma(-\ell'-s+1)}\,,\nn\\
\beta^s_{in}&=&\frac{(2)^{-\ell'+s}(-i\omega)^{-\ell'+s}\Gamma(2\ell')}{\Gamma(\ell'+s)}\,,~~~~~~~~~~~~\beta^s_{out}=\frac{(2i\omega)^{-\ell'-s}\Gamma(2\ell')}{\Gamma(\ell'-s)}\,.\label{ab}
\eea

\subsection{Near-far matching}
Matching the far region solution at $r\to0$ to the near region solution at $r\gg M$ \eqref{Rnsi}, one find the relation
\be\label{fnrs}
A^s_{far}=A^s_{near}\,,~~~~B^s_{far}=(r_+-r_-)^{-2\ell'+1}B^s_{near}\,.
\ee
For the gravitational perturbation, the $s=2$ case is for the ingoing radiative part while the $s=-2$ is for outgoing part. At infinity, the radial solutions have both outgoing and ingoing parts. One can extract the outgoing and ingoing amplitudes by the following linear combination
\bea
Z^s_{out}&\propto&A^s_{far}\alpha^{-s}_{out}+B^s_{far}\beta^{-s}_{out}\,,\\
Z^s_{in}&\propto&A^s_{far}\alpha^s_{in}+B^s_{far}\beta^s_{in}\,,
\eea
where we flip the sign of $s$ in the outgoing asymptotic coefficients since the spin components are defined according to the ingoing direction in \eqref{Rns}. The corresponding perturbation of the outgoing and ingoing Weyl scalars at infinity then take the form
\bea
\Psi_{out}(r\to\infty)&=&\sum_{\ell. m}e^{-i\omega t+im\phi}S^s_{\ell}(\theta)Z^s_{out}r^{-1+2s}e^{i\omega r}\,,\\
\Psi_{in}(r\to\infty)&=&\sum_{\ell. m}e^{-i\omega t+im\phi}S^s_{\ell}(\theta)Z^s_{in}r^{-1}e^{i\omega r}\,.
\eea
The absorption probability is the net incoming flux to the total incoming flux ratio. In the gravitational case, the energy flux at infinity is proportional to $|\Psi^s|^2$~\cite{Teukolsky:1973ha}. In this sense, the absorption probability of the gravitational radiation can be written as
\bea
\sigma^{s, bulk}_{abs}&=&1-\frac{|Z^s_{out}|^2}{|Z^s_{in}|^2}=1-\frac{|A^s_{far}\alpha^{-s}_{out}+B^s_{far}\beta^{-s}_{out}|^2}{|A^s_{far}\alpha^s_{in}+B^s_{far}\beta^s_{in}|^2}\nn\\
&=&\frac{2(A^s_{far}B^{*s}_{far}-B^s_{far}A^{*s}_{far})\left(\frac{\alpha^s_{in}}{\beta^s_{in}}\right)}{|B^s_{far}|^2}\,,
\eea
where in the last line, we invoke equations \eqref{ab} and use the fact that in the low frequency limit, the coefficient $B^s_{far}\beta^s_{in}$ is much larger than $A^s_{far}\alpha^s_{in}$ when $\mathrm{Re}(\ell')>1/2$. Substituting equations \eqref{fnrs}, \eqref{Ans}, \eqref{Bns} and \eqref{ab} into the above formula, and take $s=2$ for the gravitational case, we have
\bea\label{abssKerr}
\sigma^{s, bulk}_{abs}&=&(2\omega)^{2\ell'-1}(r_+-r_-)^{2\ell'-1}\frac{\Gamma(\ell'+2)\Gamma(-2\ell'+2)(\ell'-2)(\ell'-1)\ell'(\ell'+1)}{\Gamma(-\ell'+3)\Gamma(2\ell')\Gamma(2\ell'-1)^2\Gamma(\ell'+1/2)\Gamma(-\ell'+3/2)}\nn\\
&\times&\sinh\left(\pi\frac{4Mr_+\omega-2am}{r_+-r_-}\right)|\Gamma(\ell'-s-i2M\omega)|^2\left|\Gamma\left(\ell'-i\frac{4M^2\omega-2am}{r_+-r_-}\right)\right|^2\nn\\
&+&\mathcal{O}(\omega^{2\ell'-1})\,.
\eea

\subsection{Dual explanations in WCFT}
For the higher spin case, the form of the frequency dependence in the bulk retarded Green's function and the absorption probability are expressed in \eqref{GRsKerr} and \eqref{abssKerr}, where there are possible different normalization factors for different spin weights. Note that in the WCFT, the charge $Q$ is held fixed, invoking the relations \eqref{dl} and \eqref{OQ}, these scattering quantities \eqref{GRsKerr} and \eqref{abssKerr} match the WCFT thermal retarded correlator \eqref{GRWCFT} and the absorption cross section \eqref{absWCFT} in the $\Omega$ dependence, respectively.

However, the purely $\omega$ dependence and the spin weight parameters in the bulk scattering quantities are missing in the WCFT calculations. This is partially due to the non-locality of the WCFT operator correlations in the $Y$ coordinate, which exhibit a plane wave form~\cite{Song:2017czq}, so the charge dependence in the frequency space is a delta function with fixed value. The WCFT is a non-relativistic filed theory, a similar notion of spinning operator as in CFT is absent as well. Therefore, if the WCFT features the holographic dual of the Kerr black hole, one need to incorporate charge-dependent normalization factors of the corresponding operators to resemble the local interactions in a 2D CFT.

\section{Summary and discussion}\label{sec5}
In this paper, we consider the scalar and higher spin radiations on a generic non-extreme Kerr black hole background. We calculate the retarded Green's function and the absorption probability for the radiations, and find their matching to the thermal retarded correlator and the absorption cross section in the dual WCFT, respectively.

Unlike CFT, the WCFT has one Virasoro and one $U(1)$ Kac-Moody algebra determining the underling symmetry. This make it a good candidate as the holographic dual for a class of geometries with $SL(2, R)$ and $U(1)$ isometry. The extreme Kerr black hole enjoys such isometry in its near horizon scaling region. For non-extreme Kerr black hole, the warped conformal symmetry is not manifest in the geometry, but it can be obtained in the covariant phase space. In this sense, the WCFT could also be relevant to the holographic description of the Kerr black hole, and the present work provides further evidence on it. In the WCFT, the primary operators' momentums are translated by the energy eigenmode parameters of the bulk fields, which are inspired by the identifications between the hidden conformal generators and the global warped conformal generators. The frequency dependent $SL(2, R)$ conformal weight of the bulk scalar field is set equal to the scaling dimension of the primary operator. This leads to a charge dependent scaling dimension. However, the $U(1)$ symmetry requires the charge to be fixed on the field theory side, so the bulk field correspondingly have given fixed frequency. The matching between the bulk scattering amplitudes and the WCFT thermal correlators are then realized in the angular momentum space given a charge dependent normalization factor. For the higher spin case, the matching in the angular momentum space also exist, but the spin weight parameters are missing on the WCFT side. How to introduce additional quantum numbers in the WCFT to acquire such interpretations is an interesting next step.

\section*{Acknowledgement}
We are grateful to Jianxin Lu, Jun Nian, and Wei Song for helpful discussions. This work is supported by the NSFC Grant No. 12105045.


\begin{thebibliography}{}

\bibitem{Guica:2008mu}
M.~Guica, T.~Hartman, W.~Song and A.~Strominger,
``The Kerr/CFT Correspondence,''
Phys. Rev. D \textbf{80}, 124008 (2009)
[arXiv:0809.4266 [hep-th]].



\bibitem{Bardeen:1999px}
J.~M.~Bardeen and G.~T.~Horowitz,
``The Extreme Kerr throat geometry: A Vacuum analog of AdS(2) x S**2,''
Phys. Rev. D \textbf{60}, 104030 (1999)
[arXiv:hep-th/9905099 [hep-th]].


\bibitem{Matsuo:2009sj}
Y.~Matsuo, T.~Tsukioka and C.~M.~Yoo,
``Another Realization of Kerr/CFT Correspondence,''
Nucl. Phys. B \textbf{825}, 231-241 (2010)
[arXiv:0907.0303 [hep-th]].


\bibitem{Chen:2011wt}
B.~Chen, B.~Ning and J.~J.~Zhang,
``Boundary Conditions for NHEK through Effective Action Approach,''
Chin. Phys. Lett. \textbf{29}, 041101 (2012)
[arXiv:1105.2878 [hep-th]].




\bibitem{Bredberg:2009pv}
I.~Bredberg, T.~Hartman, W.~Song and A.~Strominger,
``Black Hole Superradiance From Kerr/CFT,''
JHEP \textbf{04}, 019 (2010)
[arXiv:0907.3477 [hep-th]].



\bibitem{Castro:2010fd}
A.~Castro, A.~Maloney and A.~Strominger,
``Hidden Conformal Symmetry of the Kerr Black Hole,''
Phys. Rev. D \textbf{82}, 024008 (2010)
[arXiv:1004.0996 [hep-th]].



\bibitem{Haco:2018ske}
S.~Haco, S.~W.~Hawking, M.~J.~Perry and A.~Strominger,
``Black Hole Entropy and Soft Hair,''
JHEP \textbf{12}, 098 (2018)
[arXiv:1810.01847 [hep-th]].



\bibitem{Nian:2023dng}
J.~Nian and W.~Tian,
``Gravitational Waves of Non-Extremal Kerr Black Holes from Conformal Symmetry,''
[arXiv:2308.03577 [hep-th]].


\bibitem{Hofman:2011zj}
D.~M.~Hofman and A.~Strominger,
``Chiral Scale and Conformal Invariance in 2D Quantum Field Theory,''
Phys. Rev. Lett. \textbf{107}, 161601 (2011)
[arXiv:1107.2917 [hep-th]].


\bibitem{Detournay:2012pc}
S.~Detournay, T.~Hartman and D.~M.~Hofman,
``Warped Conformal Field Theory,''
Phys. Rev. D \textbf{86}, 124018 (2012)
[arXiv:1210.0539 [hep-th]].



\bibitem{Compere:2013aya}
G.~Comp\`ere, W.~Song and A.~Strominger,
``Chiral Liouville Gravity,''
JHEP \textbf{05}, 154 (2013)
[arXiv:1303.2660 [hep-th]].


\bibitem{Hofman:2014loa}
D.~M.~Hofman and B.~Rollier,
``Warped Conformal Field Theory as Lower Spin Gravity,''
Nucl. Phys. B \textbf{897}, 1-38 (2015)
[arXiv:1411.0672 [hep-th]].


\bibitem{Castro:2015uaa}
A.~Castro, D.~M.~Hofman and G.~S\'arosi,
``Warped Weyl fermion partition functions,''
JHEP \textbf{11}, 129 (2015)
[arXiv:1508.06302 [hep-th]].


\bibitem{Jensen:2017tnb}
K.~Jensen,
``Locality and anomalies in warped conformal field theory,''
JHEP \textbf{12}, 111 (2017)
[arXiv:1710.11626 [hep-th]].

\bibitem{Chaturvedi:2018uov}
P.~Chaturvedi, Y.~Gu, W.~Song and B.~Yu,
``A note on the complex SYK model and warped CFTs,''
JHEP \textbf{12}, 101 (2018)
[arXiv:1808.08062 [hep-th]].


\bibitem{Song:2017czq}
W.~Song and J.~Xu,
``Correlation Functions of Warped CFT,''
JHEP \textbf{04}, 067 (2018)
[arXiv:1706.07621 [hep-th]].



\bibitem{Castro:2015csg}
A.~Castro, D.~M.~Hofman and N.~Iqbal,
``Entanglement Entropy in Warped Conformal Field Theories,''
JHEP \textbf{02}, 033 (2016)
[arXiv:1511.00707 [hep-th]].

\bibitem{Apolo:2018oqv}
L.~Apolo, S.~He, W.~Song, J.~Xu and J.~Zheng,
``Entanglement and chaos in warped conformal field theories,''
JHEP \textbf{04}, 009 (2019)
[arXiv:1812.10456 [hep-th]].



\bibitem{Anninos:2008fx}
D.~Anninos, W.~Li, M.~Padi, W.~Song and A.~Strominger,
``Warped AdS(3) Black Holes,''
JHEP \textbf{03}, 130 (2009)
[arXiv:0807.3040 [hep-th]].


\bibitem{Compere:2008cv}
G.~Comp\`ere and S.~Detournay,
``Semi-classical central charge in topologically massive gravity,''
Class. Quant. Grav. \textbf{26}, 012001 (2009)
[erratum: Class. Quant. Grav. \textbf{26}, 139801 (2009)]
[arXiv:0808.1911 [hep-th]].


\bibitem{Compere:2009zj}
G.~Comp\`ere and S.~Detournay,
``Boundary conditions for spacelike and timelike warped $AdS_{3}$ spaces in topologically massive gravity,''
JHEP \textbf{08}, 092 (2009)
[arXiv:0906.1243 [hep-th]].

\bibitem{Blagojevic:2009ek}
M.~Blagojevic and B.~Cvetkovic,
``Asymptotic structure of topologically massive gravity in spacelike stretched AdS sector,''
JHEP \textbf{09}, 006 (2009)
[arXiv:0907.0950 [gr-qc]].


\bibitem{Anninos:2010pm}
D.~Anninos, G.~Comp\`ere, S.~de Buyl, S.~Detournay and M.~Guica,
``The Curious Case of Null Warped Space,''
JHEP \textbf{11}, 119 (2010)
[arXiv:1005.4072 [hep-th]].


\bibitem{Anninos:2011vd}
D.~Anninos, S.~de Buyl and S.~Detournay,
``Holography For a De Sitter-Esque Geometry,''
JHEP \textbf{05}, 003 (2011)
[arXiv:1102.3178 [hep-th]].


\bibitem{Henneaux:2011hv}
M.~Henneaux, C.~Martinez and R.~Troncoso,
``Asymptotically warped anti-de Sitter spacetimes in topologically massive gravity,''
Phys. Rev. D \textbf{84}, 124016 (2011)
[arXiv:1108.2841 [hep-th]].



\bibitem{Compere:2013bya}
G.~Comp\`ere, W.~Song and A.~Strominger,
``New Boundary Conditions for AdS3,''
JHEP \textbf{05}, 152 (2013)
[arXiv:1303.2662 [hep-th]].


\bibitem{Song:2016gtd}
W.~Song, Q.~Wen and J.~Xu,
``Modifications to Holographic Entanglement Entropy in Warped CFT,''
JHEP \textbf{02}, 067 (2017)
[arXiv:1610.00727 [hep-th]].


\bibitem{Wen:2018mev}
Q.~Wen,
``Towards the generalized gravitational entropy for spacetimes with non-Lorentz invariant duals,''
JHEP \textbf{01}, 220 (2019)
[arXiv:1810.11756 [hep-th]].



\bibitem{Gao:2019vcc}
B.~Gao and J.~Xu,
``Holographic entanglement entropy in AdS3/WCFT,''
Phys. Lett. B \textbf{822}, 136647 (2021)
[arXiv:1912.00562 [hep-th]].


\bibitem{Apolo:2020bld}
L.~Apolo, H.~Jiang, W.~Song and Y.~Zhong,
``Swing surfaces and holographic entanglement beyond AdS/CFT,''
JHEP \textbf{12}, 064 (2020)
[arXiv:2006.10740 [hep-th]].


\bibitem{Apolo:2020qjm}
L.~Apolo, H.~Jiang, W.~Song and Y.~Zhong,
``Modular Hamiltonians in flat holography and (W)AdS/WCFT,''
JHEP \textbf{09}, 033 (2020)
[arXiv:2006.10741 [hep-th]].


\bibitem{Detournay:2020vrd}
S.~Detournay, D.~Grumiller, M.~Riegler and Q.~Vandermiers,
``Uniformization of Entanglement Entropy in Holographic Warped Conformal Field Theories,''
[arXiv:2006.16167 [hep-th]].


\bibitem{Chen:2019xpb}
B.~Chen, P.~X.~Hao and W.~Song,
``R\'enyi mutual information in holographic warped CFTs,''
JHEP \textbf{10}, 037 (2019)
[arXiv:1904.01876 [hep-th]].



\bibitem{Chen:2022fte}
B.~Chen, Y.~Liu and B.~Yu,
``Reflected entropy in AdS$_{3}$/WCFT,''
JHEP \textbf{12}, 008 (2022)
[arXiv:2205.05582 [hep-th]].



\bibitem{Song:2019txa}
W.~Song and J.~Xu,
``Structure Constants from Modularity in Warped CFT,''
JHEP \textbf{10}, 211 (2019)
[arXiv:1903.01346 [hep-th]].



\bibitem{Azeyanagi:2018har}
T.~Azeyanagi, S.~Detournay and M.~Riegler,
``Warped Black Holes in Lower-Spin Gravity,''
Phys. Rev. D \textbf{99}, no.2, 026013 (2019)
[arXiv:1801.07263 [hep-th]].




\bibitem{Aggarwal:2019iay}
A.~Aggarwal, A.~Castro and S.~Detournay,
``Warped Symmetries of the Kerr Black Hole,''
JHEP \textbf{01}, 016 (2020)
[arXiv:1909.03137 [hep-th]].

\bibitem{Zeldovich:1971}
Y.~B.~Zel'dovich,
``Generation of waves by a rotating body,''
Pis'ma Zh. Eksp. Teor. Fiz. \textbf{14}, 270 (1971)
[JETP Lett. \textbf{14}, 180 (1971)].

\bibitem{Zeldovich:1972}
Y.~B.~Zel'dovich,
``Amplifcation of cylindrical electromagnetic waves from a rotating body,''
Zh. Eksp. Teor. Fiz \textbf{62}, 2076 (1972)
[Sov. Phys. JETP \textbf{35}, 1085 (1972)].




\bibitem{Maldacena:1997ih}
J.~M.~Maldacena and A.~Strominger,
``Universal low-energy dynamics for rotating black holes,''
Phys. Rev. D \textbf{56}, 4975-4983 (1997)
[arXiv:hep-th/9702015 [hep-th]].


\bibitem{Gubser:1997cm}
S.~S.~Gubser,
``Absorption of photons and fermions by black holes in four-dimensions,''
Phys. Rev. D \textbf{56}, 7854-7868 (1997)
[arXiv:hep-th/9706100 [hep-th]].


\bibitem{Teukolsky:1972my}
S.~A.~Teukolsky,
``Rotating black holes - separable wave equations for gravitational and electromagnetic perturbations,''
Phys. Rev. Lett. \textbf{29}, 1114-1118 (1972)


\bibitem{Teukolsky:1973ha}
S.~A.~Teukolsky,
``Perturbations of a rotating black hole. 1. Fundamental equations for gravitational electromagnetic and neutrino field perturbations,''
Astrophys. J. \textbf{185}, 635-647 (1973)






\bibitem{Bonelli:2021uvf}
G.~Bonelli, C.~Iossa, D.~P.~Lichtig and A.~Tanzini,
``Exact solution of Kerr black hole perturbations via CFT2 and instanton counting: Greybody factor, quasinormal modes, and Love numbers,''
Phys. Rev. D \textbf{105}, no.4, 044047 (2022)
[arXiv:2105.04483 [hep-th]].




\bibitem{Goldberg:1967}
J.~N.~Goldberg, A.~J.~Macfarlane, E.~T.~Newman, F.~Rohrlich, E.~C.~G.~Sudarshan,
``Spin-s Spherical Harmonics and edth,''
J. Math. Phys. \textbf{8}, 2155-2161 (1967)









\end{thebibliography}
\end{document}